# Faculty citation measures are highly correlated with peer assessment of computer science doctoral programs[1]

**Slobodan Vucetic, Ashis Kumar Chanda, Shanshan Zhang, Tian Bai, Aniruddha Maiti**
Department of Computer and Information Sciences, Temple University

## ABSTRACT

We study relationship between peer assessment of quality of U.S. Computer Science (CS) doctoral programs and objective measures of research strength of those programs. In Fall 2016 we collected Google Scholar citation data for 4,352 tenure-track CS faculty from 173 U.S. universities. The citations are measured by the *t10* index, which represents the number of citations received by the 10th highest cited paper of a faculty. To measure the research strength of a CS doctoral program we use 2 groups of citation measures. The first group of measures averages *t10* of faculty in a program. Pearson correlation of those measures with the peer assessment of U.S. CS doctoral programs published by the U.S. News in 2014 is as high as 0.890. The second group of measures counts the number of well cited faculty in a program. Pearson correlation of those measures with the peer assessment is as high as 0.909. By combining those two groups of measures using linear regression, we create the Scholar score whose Pearson correlation with the peer assessment is 0.933 and which explains 87.2% of the variance in the peer assessment. Our evaluation shows that the highest 62 ranked CS doctoral programs by the U.S. News peer assessment are much higher correlated with the Scholar score than the next 57 ranked programs, indicating the deficiencies of peer assessment of less-known CS programs. Our results also indicate that university reputation might have a sizeable impact on peer assessment of CS doctoral programs. To promote transparency, the raw data and the codes used in this study are made available to research community at http://www.dabi.temple.edu/~vucetic/CSranking/.

## 1 INTRODUCTION

Various published rankings of universities and specialized academic programs have become a major influencer for students deciding which university to attend, faculty deciding where to work, government bodies deciding where and how to invest their education and research funding, and the university leaders deciding how to grow their institutions [10]. There is a general agreement that quality of a university or a program depends on many factors and that different ranking metrics might be appropriate for different types of users. However, there are major points of contention when it comes to agreeing on ranking methodology, such as which measures to use for ranking, how to combine them, and whether and how to customize ranking depending on the purpose [22]. In addition to methodological questions, there is also an issue of transparency. The most popular rankings, such as the U.S. News ranking, the Academic Ranking of World Universities (ARWU), or the QS World University Rankings, typically do not provide access to raw data, the concrete algorithms for processing the raw data, and a precise justification for the metrics being used. Given the impact of rankings, there is a need to better understand properties and relationships between various quality measures and come up with justifiable, transparent, and interpretable rankings that promote education and research quality [12]. The objective of this study is to contribute towards achieving such an objective. We focus on ranking of the U.S. doctoral programs in computer science (CS), but our insights and results should be useful to ranking of other disciplines.

Quality measures can be broadly grouped into objective measures, such as average research funding per faculty, and subjective, such as peer assessment. Rankings of graduate programs explained next are representative of the ways quality measures have been used. Arguably the most influential ranking of the CS doctoral programs in the U.S. is the one provided by the U.S. News[2]. This ranking is based purely on the peer assessment. More specifically, the most recent U.S. News ranking published in 2014 is based on surveys from 2009 and 2013 sent to the chairs of the CS departments and graduate program directors, in which the participants were asked to rank 177 Ph.D.-granting CS programs in the U.S. on a scale from 1 to 5, 1 meaning "marginal" and 5 meaning "outstanding", or enter "do not know" if not sufficiently familiar with the program. The average score of the 2009 and 2013 surveys for a program is reported as the final score. Raw and processed data are not published and no information is provided about potential issues with the collected data beyond the fact that the survey response rate was 35%. An open question with the peer assessments is whether the peers know enough about a sufficient number of other programs to provide a fair and accurate assessment.

---

[1] A version of this material has been submitted to Communications of the ACM for consideration for publication
[2] http://grad-schools.usnews.rankingsandreviews.com/best-graduate-schools/top-science-schools/computer-science-rankings

Unlike the U.S. News ranking, the 2015 ARWU [18] ranking[3] is based on objective measures. The measures include counts of papers published in computer science venues, fraction of papers published in selected computer science journals, number of highly cited faculty, and numbers of Turing Award winners. The final ranking is a weighted average of those measures. The choice of weights is not justified and this was met with some critique in the scientometrics community [5, 7].

U.S. News ranking of doctoral programs in engineering[4] uses a weighted average of objective measures (accounting for 60% of the score), such as research expenditures, student selectivity, and doctoral degrees awarded, and subjective measures (accounting for 40% of the score), such as peer assessment and recruiter assessment. Similarly to the ARWU, justification for the ranking formula is not provided.

Ranking of CS doctoral programs published in 2010 by the U.S. National Research Council (NRC) [2] is notable for its effort to provide a justifiable ranking formula. The NRC collected objective measures such as the number of publications per faculty, fraction of the faculty supported by grants, number of grants, diversity of the faculty and students, student GRE scores, graduate student funding, number of doctoral students, and time to degree. Many of those measures were obtained directly from the departments. The NRC also surveyed faculty to assess the peer institutions on different measures of perceived quality. Instead of coming up with an arbitrary metric to combine objective and subjective measures, the NRC ranking group fitted a regression model that predicted subjective measures based on the objective measures. The resulting regression model was used to provide the ranking. Unfortunately, the NRC ranking project had issues with data quality [4] and the published ranking did not have good reception[5].

We find the NRC idea of calculating the ranking function through regression modeling better justified than the alternatives, such as relying solely on peer assessment or on an arbitrary combination of objective and subjective measures. In this study, we address the pitfalls of the NRC ranking project and demonstrate that regression analysis is a viable approach for ranking of graduate programs. To address the data quality issue, we make sure that our objective measures are unbiased and with high coverage. Among many possible objective measures, we focus solely on citation measures of CS faculty obtained from Google Scholar,[6] a free web service with an excellent coverage of research papers and their citations. Consistent with scientometrics research [23], our assumption is that citations of papers are indicative of research impact of faculty and that objective measures derived from citations are a good predictor of quality of doctoral programs. Google Scholar is particularly appropriate for our study due to its coverage of conference papers, which computer scientists traditionally value as highly as journal papers.

As our results show, it is possible to come up with a simple formula based on citation measures that is highly correlated with the peer assessment of CS doctoral programs. Our study provides an insight into the quality of peer assessments, correlation between citation measures and the peer assessments, and an impact of university reputation on peer assessments. We publish all the raw data used in this study, all the code used to produce the results shown in this paper, and the resulting ranking. We argue that the resulting ranking formula is an informative, simple, and transparent measure of quality of CS doctoral programs in the U.S. We also discuss limitations of the current study and propose avenues for further research on the topic of rankings of academic programs.

---

[3] http://www.shanghairanking.com/SubjectCS2015.html
[4] https://www.usnews.com/education/best-graduate-schools/articles/engineering-schools-methodology?int=9d0e08
[5] http://www.chronicle.com/article/Too-Big-to-Fail/127212/
[6] https://scholar.google.com/

## 2 DATA AND METHODS

In this section we explain the data collection process, which took place during Fall of 2016. Our data collection team had 3 undergraduate students, 5 computer science graduate students, and a computer science professor.

### 2.1 U.S. News (USN) Data

U.S. News (USN) provides a well-known ranking of graduate programs in the United States. We downloaded the most recent USN ranking of 173 U.S. Computer Science (CS) doctoral programs published in 2014. As explained in the introduction, the USN ranking of CS doctoral programs is based purely on peer assessment[7]. Each program receiving at least 10 ratings from its peers obtains a score equal to the average of the received scores. CS doctoral programs with the average score of at least 2.0 are ranked and their score is published. The remaining programs are not ranked. No information is available about any potential source of systematic bias in the survey responses or any effort to deal with the missing data and outliers. Raw survey data are not available. In the remainder of the paper, we refer to the 2014 USN peer assessment scores of CS doctoral programs as the *USN CS scores*.

We also downloaded the 2017 USN National University Ranking,[8] which evaluates the quality of undergraduate programs at U.S. universities. Each university in this ranking is assigned a score between 0 and 100 based on the numerous measures of quality[9]. We refer to those scores as the *USN university scores*.

### 2.2 CS Faculty Name Data

We manually collected names of 4,728 tenure-track CS professors from the 173 programs ranked by USN. To be counted as a CS professor, a tenure-track faculty had to be claimed by a web site of a computer science department or college. To find a web site of a CS department, we followed a web link provided by the U.S. News. In addition, we performed Google search for queries such as "<university_name> computer science" or "<university_name> cs". In most cases, the web sites of CS departments have easily accessible pages named "People", "Faculty", or "Directory", containing lists of all of the CS faculty and their appointments. In a few cases, people pages do not specify appointments and we had to access the individual faculty pages to extract this information.

In a smaller number of universities, CS faculty are part of joint departments such as "Electrical Engineering and Computer Sciences" or "Computer Science and Engineering", making it harder to identify the CS faculty from people pages. Selecting the CS faculty in such departments is challenging because the boundaries between the disciplines are not always clear. The faculty that are the hardest to distinguish are those in the areas of computer engineering, signal processing, information theory, and control. We decided to err on the generous side and counted the CS faculty as all faculty who have at least some publications in the CS venues.

Several universities have schools or colleges of computing or informatics, which contain a mix of CS faculty and faculty from areas such as library science, informatics, information sciences, or management information systems. In cases where the CS faculty are not clearly distinguished on people pages, we again decided to count the CS faculty as all faculty with at least some publications in the CS venues. In two cases (Georgia Tech, NYU), the universities have multiple CS departments in the same or different colleges, and we counted all faculty from those departments as CS faculty.

Another issue is dealing with affiliated faculty or faculty with secondary or joint appointments in the CS departments. In many cases, people pages clearly separate such faculty from those with the primary appointments. In such cases, we only include the primary appointments to our list. In other cases, the affiliated and primary faculty are mixed together on the people pages. The type of an appointment is sometimes clearly stated, sometimes it can only become evident after studying the individual faculty pages, and sometimes the information is not available on the web. When people pages mixed primary and non-primary appointments in the same list, we again decided to err on the generous side and included all listed tenure-track faculty, unless it was clearly specified that they are affiliated, adjunct, courtesy, or visiting faculty.

[7] https://www.usnews.com/education/best-graduate-schools/articles/science-schools-methodology
[8] https://www.usnews.com/best-colleges/rankings/national-universities
[9] https://www.usnews.com/education/best-colleges/articles/how-us-news-calculated-the-rankings

On the flip side of the affiliated faculty issue, in many universities there are faculty who actively or mostly publish in the CS venues, but whose primary appointments are in departments such as electrical engineering, computer engineering, biology, or statistics, or in schools of business or public health. Quite often, such faculty are not mentioned on people pages of the CS departments. We were not able to devise an unbiased and cost effective way to identify such faculty, and decided to consider only faculty explicitly mentioned on people pages of the CS departments for inclusion to our list.

Overall, we collected the names of 4,728 tenure-track faculty, 1,114 of whom are assistant professors, 1,271 associate professors, and 2,343 full professors. For the sake of transparency, the collected faculty names are available on http://www.dabi.temple.edu/~vucetic/CSranking/.

Computer science departments are young: 23.6% (1,114/4,728) of their faculty are assistant professors and in 35 of the 173 programs more than 1/3 of the faculty are assistant professors. Since the assistant professors are in most cases only starting their academic careers and are establishing their publication record, we treat them differently from associate and full professors. In the remainder of the paper, we refer to the associate and full professors as the *senior faculty*.

## 2.3 Google Scholar Data

Of the 4,728 CS faculty, 3,359 has Google Scholar profiles (71.0% coverage) and of the 3,614 senior CS faculty, 2,453 (67.9% coverage) has Google Scholar profiles. A Google Scholar profile lists all publications of a faculty, with citation counts for each paper and aggregate citation measures such as the total citation counts, *h*-index, and *i10*-index. The popular *h*-index [13] is the highest integer $x$ for which it can be stated that the author published $x$ papers that are cited at least $x$ times. The *i10*-index provides a count of papers cited 10 times or more.

One option was to perform our study using only the citation data of the faculty with Google Scholar profiles. However, we observed that less cited faculty are less likely to have the profile and that the citation data obtained from the 3,359 profiles is a biased sample from the 4,728 CS faculty. To prevent a negative impact of bias on the analysis, we proceeded with collecting an unbiased sample of faculty citation data. To achieve this in a limited amount of time, we introduced a new citation measure, as explained next.

## 2.4 *t10* Index

We define the *t10* index as the number of citations of the faculty's 10th highest-cited paper. The new index is more convenient than *h*-index because it is faster to obtain by manual search. For example, rather than having to find the 50 highest cited papers authored by a faculty to establish that his or her *h*-index is 50, the *t10* index could be obtained by identifying only the faculty's 10 highest cited papers.

We obtained *t10* for 4,352 of the 4,728 CS faculty (92.0% coverage) and for 3,330 of the 3,614 senior CS faculty (92.1% coverage), by manual search of Google Scholar. For faculty with the Google Scholar profile, extracting *t10* was straightforward because the papers in the profile are sorted by the citations. For faculty without the profile, the curators had to manually process the papers retrieved by Google Scholar. Given a faculty name, the top papers retrieved by the Google Scholar search engine are typically those that contain the name in the author list. Moreover, higher cited papers are typically placed above the lower ranked ones. If a faculty name is unique, the list of papers retrieved by Google Scholar is often easy to parse and *t10* could be extracted with relatively little effort. However, when a faculty has a common name or when a faculty name listed on the people pages does not precisely match the name listed in his or her papers, obtaining *t10*-index can be too time consuming. To save time, the curators were instructed to abort the extraction if it took more than 4 minutes. As a result, we did not collect *t10* for 376 of the 4,728 CS faculty (8.0%). Since a faculty's name should not have an influence on his or her citation count, the resulting sample of faculty with known *t10* can be treated as an unbiased sample of the senior CS faculty.

## 2.5 Program Strength Measures

We propose two approaches for using individual faculty citation indices to calculate the citation strength of a whole program.

**Averaged Citation Measures**. One way to measure strength of a program is to average citations of its individual faculty members [16]. We explore 3 different averaging schemes. The first is based on the median of *t10* values of senior CS faculty. We denote the median of *t10* values as *m10*. The second averaging scheme is based on the geometric mean of *t10* values of senior CS faculty. We denote the geometric mean of (1+*t10*) values as *g10*. The third averaging scheme is based on the percentiles of *t10* values of senior CS faculty. We denote the average *t10* percentile of the faculty as *p10*. We do not count assistant professors for any of the averaged measures because their citation numbers are typically smaller and their inclusion would hurt departments with large fractions of assistant professors.

**Cumulative Citation Measures**. Another way to measure strength of a program is to count the number of highly cited faculty in a program. To define a highly cited faculty, we have to decide on a *t10* threshold. All faculty above the *t10* threshold are considered highly cited. We introduce *cN*, which denotes the number of faculty whose *t10* is higher than *N* % of the senior faculty. For example, *c40* counts the number of CS faculty with *t10* larger than 40% of all senior faculty. To find *cN*, we consider faculty at all ranks, including the assistant professors.

## 2.6 Regression Models

Our preliminary results showed that combining one averaged and one cumulative citation measure could increase the correlation with the USN CS scores. These results also showed that the particular choice of a measure often does not have a major impact on the correlation strength. Finally, the results showed that linear regression with 2 measures is nearly as successful as linear regression with more than 2 measures or nonlinear regression with 2 or more measures. Thus, we use regression models that combine averaged and cumulative citation measures into an aggregated score, such that it matches the USN CS scores as closely as possible.

The regression models we consider are of type $s_i = \beta_0 + \beta_1 a_i + \beta_2 c_i$, where $s_i$ is the predicted USN CS score, $a_i$ is an aggregated citation measure, and $c_i$ is a cumulative citation measure of the *i*-th program. The regression parameters are $\beta_0$, $\beta_1$, and $\beta_2$. Instead of learning the intercept parameter $\beta_0$, we set it to $\beta_0 = 1$ by default. The primary justification is that a program with $a_i = 0$ and $s_i = 0$ does not have research-active faculty and, based on the peer assessment instructions by USN, such a program would be scored as 1 ('Marginal'). The secondary justification is that the resulting regression models are simpler because they require fitting only 2 regression parameters, $\beta_1$, and $\beta_2$.

Since in Section 3.5 we proposed 3 averaged program measures and a range of cumulative measures *cN*, for $0 < N < 100$, we train one regression model for each combination of the averaged and the cumulative measure. We also create an aggregate score, by averaging the individual regression models.

# 3 RESULTS

## 3.1 Basic Statistics

**U.S. News ranking**. Of the 173 CS doctoral programs ranked in 2014, USN assigns scores of 2.0 or higher to 119 programs, while 54 programs have scores below 2.0 that remain unpublished. Seventeen programs have scores of 4.0 or higher and 4 programs (CMU, MIT, Stanford, Berkeley) have the maximum possible score of 5.0. We compare the USN CS scores with the 2017 USN university scores. The Pearson's correlation coefficient between the two scores for 113 programs covered by both rankings is relatively high at 0.681.

**Faculty data**. **Fig. 1** shows the distribution of the department sizes, defined as the number of tenure-track faculty in each of the 173 ranked CS programs. The distribution is quite wide, with the median faculty size 22, the mode 15, the minimum size 4, and the maximum size 143 (CMU). The correlation coefficient between the department sizes and the USN CS scores of the 119 ranked CS programs is 0.676, indicating that larger departments are more likely to be higher ranked.

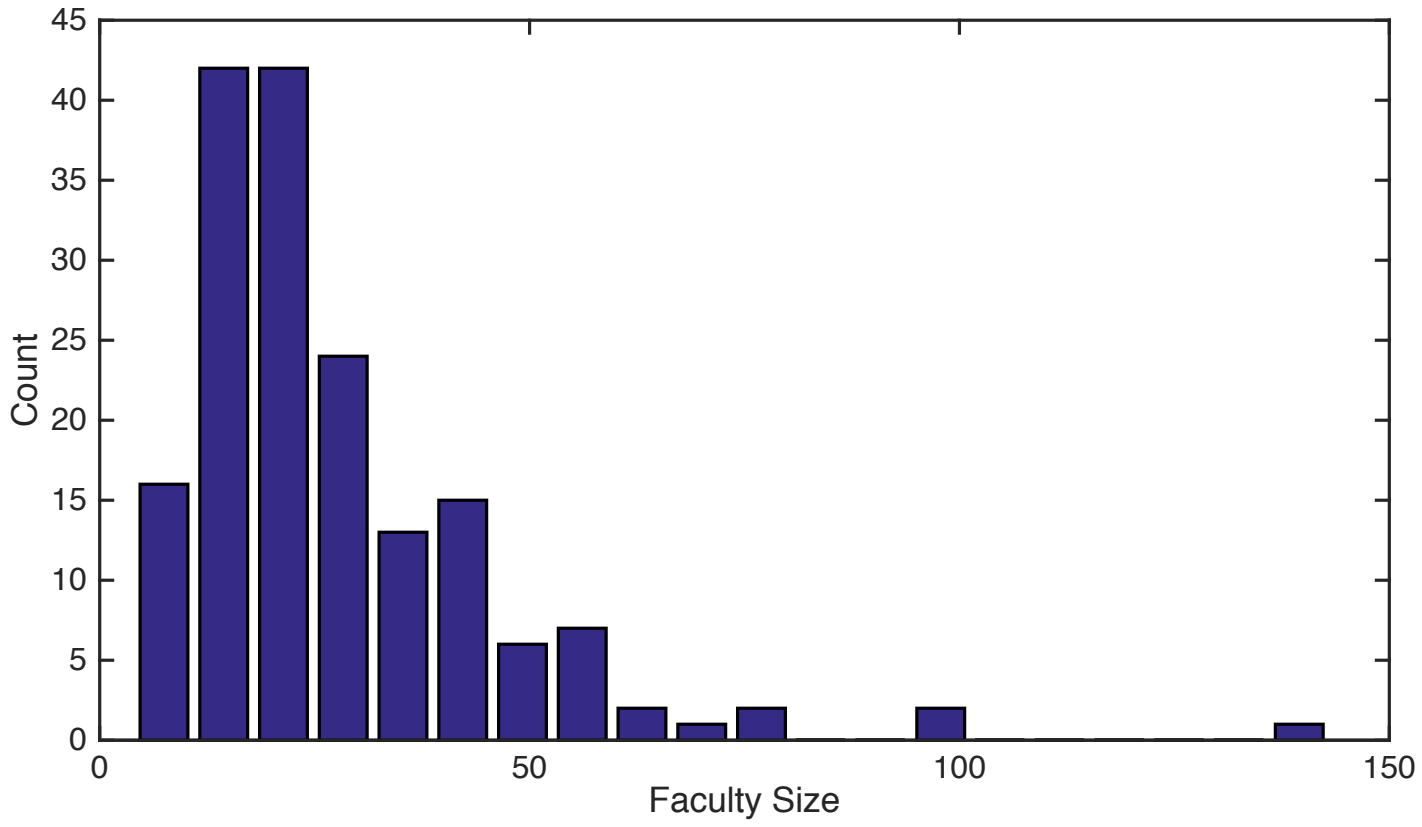

**Figure 1:** The distribution of the U.S. CS department sizes

## 3.2 Distribution of the Citation Indices

In **Fig. 2** we show the histogram of *t10* for the 3,330 senior CS faculty. Since *t10* resembles lognormal distribution, the histogram of *t10* is shown in a log scale. We can observe a bump at low values, which represents 89 senior faculty with *t10* = 0, who have less than 10 cited papers listed in Google Scholar. The median of *t10* is 89 and the percentiles of *t10* are shown in **Table 1**. For example, to be in the 90th percentile of all senior CS faculty in the U.S., a faculty should have 10 papers cited at least 370 times.

The correlation coefficient between logarithms of *h*-index and *t10* for the 2,453 senior CS faculty having both indices is 0.937. The correlation is sufficiently high to conclude that *t10* index is a good proxy for *h*-index.

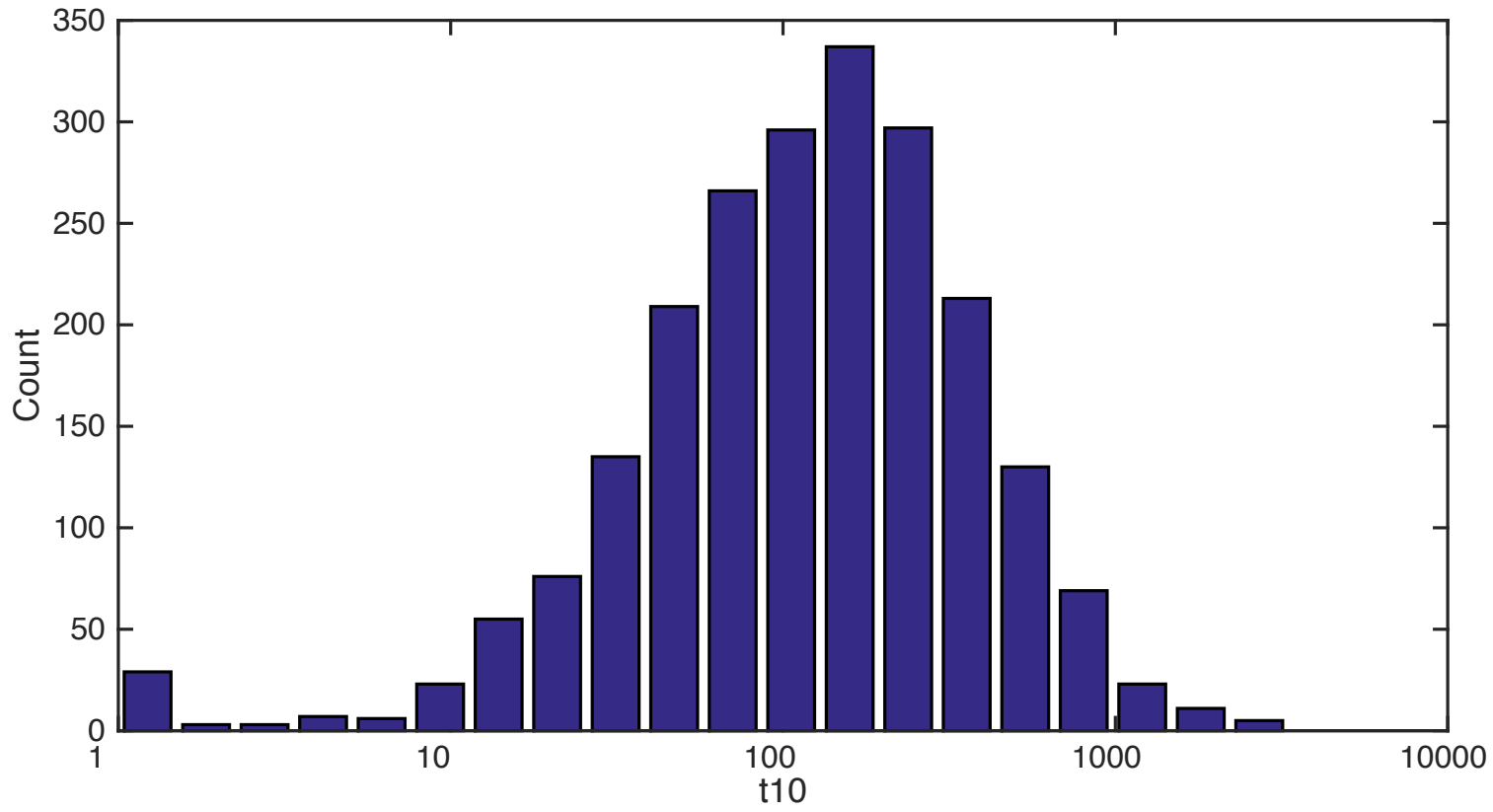

**Figure 2:** Histogram of *t10* of associate and full CS professors

**Table 1:** Percentiles of *t10*

| Percentile | t10 | Percentile | t10 |
|---|---|---|---|
| 10% | 16 | 70% | 166 |
| 20% | 31 | 80% | 231 |
| 30% | 47 | 90% | 370 |
| 40% | 65 | 95% | 543 |
| 50% | 90 | 98% | 780 |
| 60% | 123 | 99% | 1,078 |

### 3.3 Scholar Profile Bias

While the median of *t10* for the 3,330 senior CS faculty is 89, it increases to 111 among 2,453 of those faculty who also have the Google Scholar profile, and it drops to 44 among 877 of those faculty without the profile. In **Fig. 3** we show a stacked bar plot of the numbers of faculty with and without Google Scholar profile as a function of their *t10* percentile. Among the 326 faculty below the 10th percentile, 213 (65.3%) do not have the Google Scholar profile, while among the 334 faculty above the 90-th percentile only 37 (11.1%) are without the profile. These results clearly indicate that CS faculty with the Google Scholar profile are a biased sample of the entire CS faculty and validate our effort to gather the *t10* indices and use them in our study instead of the *h*-index.

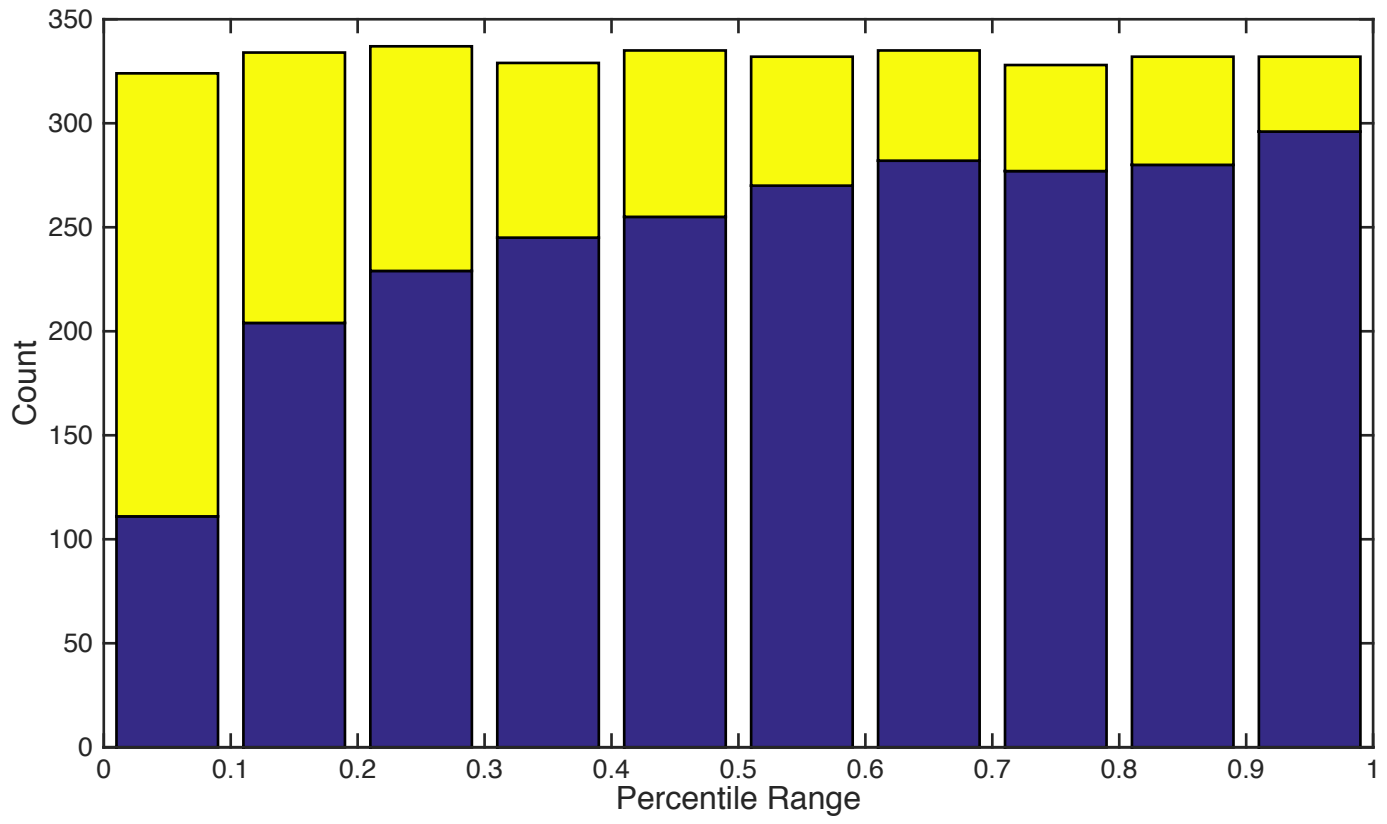

**Figure 3:** Numbers of tenured CS faculty with (blue) and without (yellow) Google Scholar profile as a function of the *t10* percentile

### 3.4 Correlation between citation measures and USN CS scores

To evaluate ability of the averaged measures to explain peer assessment of the CS programs, we calculate their correlation with the USN CS scores. The 'Original' row of **Table 2** shows Pearson correlation coefficients between the proposed 3 averaged measures (*m10*, *g10*, and *p10*) and 4 cumulative citation measures (*c20*, *c40*, *c60*, *c80*) and the USN CS scores. The values range from 0.794 to 0.882, indicating a strong correlation between the averaged citation measures and the USN CS scores.

Since the distribution of *m10* and *g10* measures is heavy-tailed, we also explore their logarithmic and square root transformation. The correlation between the transformed measures and the USN CS scores are shown in rows 'Log' and 'Sqrt' of **Table 2**. The square root transformation of *m10* results in the highest correlation (0.890), while *p10* has the highest correlation in its original form. This result supports our original hypothesis that peer assessment of the quality of CS doctoral programs is closely tied to the quality of CS faculty research, which is quantified by citation measures.

In **Table 2** we also show Pearson correlation for logarithm and square root transformation of the 4 cumulative measures. The correlation between the square root of *c60* and the USN CS score reaches 0.909, which is the highest observed correlation of all the measures we examined. This result confirms that the count of highly productive faculty is a very accurate proxy to the perceived CS program quality. Measure *c80* reaches comparably high correlation, while *c20* is the least correlated. The low correlation of *c20* is confirming that the department size is not the strongest factor in determining the strength of a CS doctoral program.

**Table 2:** Correlation of averaged program measures with USN CS scores

| Transform | *m10* | *g10* | *p10* | *c20* | *c40* | *c60* | *c80* |
|---|---|---|---|---|---|---|---|
| Original | 0.880 | 0.875 | **0.882** | 0.794 | 0.844 | 0.874 | 0.842 |
| Log | 0.865 | 0.856 | 0.870 | 0.807 | 0.840 | 0.875 | 0.904 |
| Sqrt | **0.890** | **0.887** | 0.861 | **0.825** | **0.877** | **0.909** | **0.906** |

### 3.5 Regression Analysis

By combining one of the averaged citation measures ($\sqrt{m10}$, $\sqrt{g10}$, $p10$) and one of the cumulative citation measures ($\sqrt{c40}$, $\sqrt{c60}$, $\sqrt{c80}$) to train a regression model of the type described in section 2.6, we train 9 different regression models. For training, we use 119 CS doctoral programs scored with rates 2.0 and above by the USN CS scores. In **Table 3** we show the coefficient of determination $R^2$, which shows percent explained variance, for each of the 9 regression models. In **Table 4**, we show Pearson correlation between the regression model outputs and the USN CS scores for the 9 regression models. The correlation of all 9 models is in the range from 0.920 to 0.934, which is higher than for any of the individual citation measures from **Table 2**. It can also be observed that the best 4 models (in bold in **Tables 3** and **4**) are a combination of either $\sqrt{m10}$ or $\sqrt{g10}$ averaged measure and $\sqrt{c40}$ or $\sqrt{c60}$ cumulative measure. In **Table 5** we show the parameters of those 4 models. The best overall model, which achieves $R^2 = 0.869$ and correlation coefficient 0.934, is a combination of $\sqrt{m10}$ and $\sqrt{c60}$ measures,

$$s = 1 + 0.130\sqrt{m10} + 0.218\sqrt{c60}\,.$$

Thus, if the median faculty in a CS program has $t10 = 100$ and there are 9 faculty over the 60-th percentile (with $t10 \geq 123$) based on $t10$ index, the calculated score of that program would be 2.95.

**Table 3:** $R^2$ of regression models $1 + \beta_1 a_i + \beta_2 c_i$. Rows represent different $a_i$ measures and the columns represent different $c_i$ measures

| | $\sqrt{c40}$ | $\sqrt{c60}$ | $\sqrt{c80}$ |
|---|---|---|---|
| $\sqrt{m10}$ | **0.864** | **0.869** | 0.846 |
| $\sqrt{g10}$ | **0.860** | **0.867** | 0.841 |
| $p10$ | 0.845 | 0.856 | 0.845 |

**Table 4:** Pearson correlation between regression models $1 + \beta_1 a_i + \beta_2 c_i$ and USN CS scores. Rows represent different $a_i$ measures and the columns represent different $c_i$ measures.

| | $\sqrt{c40}$ | $\sqrt{c60}$ | $\sqrt{c80}$ |
|---|---|---|---|
| $\sqrt{m10}$ | **0.931** | **0.934** | 0.923 |
| $\sqrt{g10}$ | **0.929** | **0.933** | 0.920 |
| $p10$ | 0.920 | 0.927 | 0.922 |

**Table 5:** Parameters of the four best individual ranking models and Scholar model.

| | $\sqrt{m10}$ | $\sqrt{g10}$ | $\sqrt{c40}$ | $\sqrt{c60}$ |
|---|---|---|---|---|
| **Model 1** | 0.111 | | 0.223 | |
| **Model 2** | 0.130 | | | 0.218 |
| **Model 3** | | 0.113 | 0.225 | |
| **Model 4** | | 0.133 | | 0.220 |
| **Joint Model** | 0.060 | 0.062 | 0.112 | 0.109 |

### 3.6 Scholar Model

**Tables 3** and **4** indicate that 4 of the 9 models are more accurate than others. To exploit this variability, we create a joint model by averaging scores from those 4 models. This joint model has $R^2 = 0.874$ and correlation coefficient 0.935, which makes it more accurate than any of the individual models. By averaging the parameters of the 4 individual models from **Table 5**, we can obtain the parameters of the joint model listed in the last row of the table.

In **Figure 4** we show a scatterplot between the USN and the joint model scores for the 173 CS programs. For the 55 CS doctoral programs unranked by the USN, their default USN CS score is set to 1.5 in the figure. The figure illustrates a remarkable correlation between the peer assessed USN CS scores and the objectively measured joint model scores.

A closer look at the scatterplot reveals that two groups of CS programs can be distinguished with respect to the correlation between joint model scores and USN CS scores. The first group contains 62 programs scored 2.7 and higher by the USN. The correlation between the USN CS scores and joint model scores in this group is 0.911. The second group contains 57 programs with USN CS scores between 2.0 and 2.6. The correlation between the USN CS scores and joint model scores in this low scoring group is only 0.360, showing a sizeable discrepancy between those two types of scores. Our hypothesis is that the CS programs whose USN CS scores are between 2.0 and 2.6 might not be sufficiently well known among the peers at the national level to allow objective and reliable peer assessments.

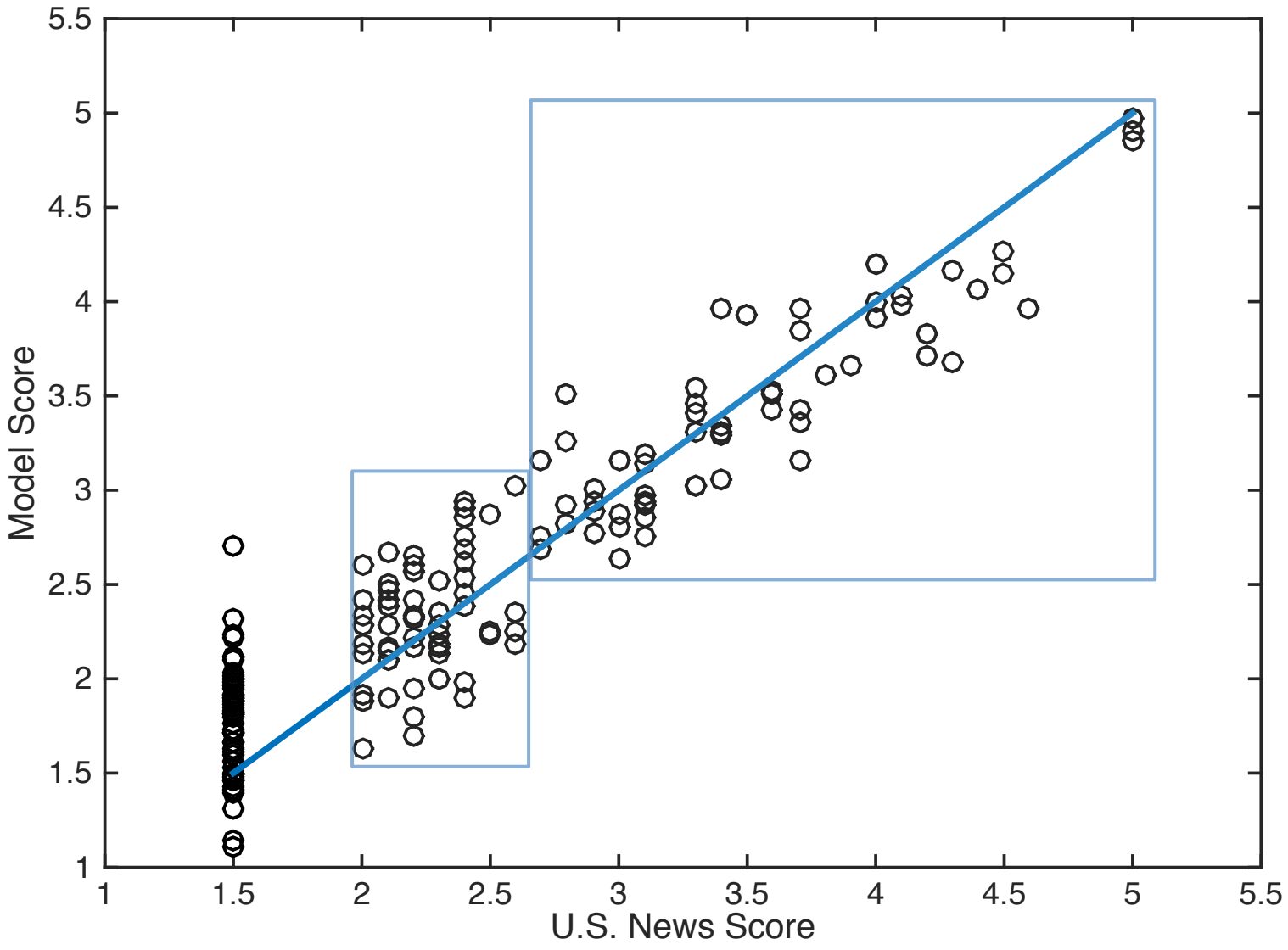

**Figure 4:** Comparison of aggregated model scores and USN CS scores of 173 CS graduate programs. For 55 programs not ranked by the USN, we set the default score of 1.5.

We hypothesized that USN CS scores of the 57 programs with the scores between 2.0 and 2.6 might be too noisy to help the regression model. To explore this, we trained another joint model using only the data from the 62 CS programs scored above 2.6 by USN CS ranking. The resulting individual models and the joint model are shown in **Table 6**. While there is a large similarity between these models and the ones trained on 119 CS programs listed in **Table 5**, the main difference is that the models from **Table 6** give a slightly higher weight to cumulative citation measures. When measured on the top 62 CS programs, the joint model from **Table 6** has $R^2$ = 0.830 and correlation 0.913, which is slightly larger than $R^2$ = 0.818 and correlation 0.911 of the joint model from **Table 5**. When measured on all 119 ranked CS programs, the joint model from **Table 6** has $R^2$ = 0.872 and correlation 0.935, which is virtually identical to the joint model from **Table 5**. We conclude that USN CS scores of programs ranked from 2.0 to 2.6 are indeed too noisy to be helpful for regression modeling. As a result, we accept the joint model from **Table 6** as the best model for ranking of CS doctoral programs. We refer to the joint model from **Table 6** as the **Scholar model** and to the outputs of this model as the **Scholar scores**. The Scholar scores are calculated as

$$s = 1 + 0.058\sqrt{m10} + 0.059\sqrt{g10} + 0.121\sqrt{c40} + 0.127\sqrt{c60}\ .$$

In **Table 7** we show rankings based on the Scholar scores and some relevant statistics for all 173 CS doctoral programs in our study.

**Table 6:** Parameters of the four best individual ranking models and Scholar model, trained on 62 top ranked CS programs by USN CS ranking.

| | $\sqrt{m10}$ | $\sqrt{g10}$ | $\sqrt{c40}$ | $\sqrt{c60}$ |
|---|---|---|---|---|
| **Model 1** | 0.111 | | 0.240 | |
| **Model 2** | 0.122 | | | 0.251 |
| **Model 3** | | 0.113 | 0.242 | |
| **Model 4** | | 0.124 | | 0.256 |
| **Scholar Model** | 0.058 | 0.059 | 0.121 | 0.127 |

## 3.7 Impact of Reputation

We are interested in the impact of the overall university reputation on the CS program ranking. We train regression models of type $s_i = 1 + \beta_1 a_i + \beta_2 c_i + \beta_3 us_i$, where $us_i$ is the score of the $i$-th university according to the 2017 USN National University ranking. We note that the maximum USN university score is 100 (Princeton) and the lowest listed score is 20. For universities that do not have a listed score, we set their score to 20 by default. We train 4 regression models that use one of the averaged measures $\sqrt{m10}$ or $\sqrt{g10}$, one of the cumulative measures $\sqrt{c40}$ or $\sqrt{c60}$, and 2017 USN university score. The training was done using 119 universities with USN CS score of 2.0 and above. By averaging outputs of these 4 models, the resulting aggregated model has $R^2$ = 0.888 and correlation coefficient 0.942, which is a modest increase in accuracy compared to the models in **Table 5**, which do not rely on USN university scores. This result indicates that the overall university reputation might have an impact on the peer assessments of the CS doctoral programs.

To get an insight into the result, let us look at the most accurate individual model whose $R^2$ = 0.884 and correlation coefficient 0.941,

$$s = 1 + 0.090\sqrt{m10} + 0.238\sqrt{c60} + 0.0061 \cdot us\ .$$

This model adds 0.61 to the score of the Princeton University CS doctoral program and only 0.12 to the CS programs from universities unranked by the 2017 USN National University ranking. As a result, if the citation measures of two CS programs are identical, their scores assigned by this model could differ by as much as 0.49 depending on their university score.

## 3.8 Qualitative Analysis of Scholar Scores

Based on the $R^2$ of the Scholar model, we see that measures derived from faculty citations collected during the Fall of 2016 can explain 87.4% of the variance of the peer-assessed USN CS scores. For the qualitative analysis, we look at the programs with the largest discrepancies between USN CS scores and Scholar scores (see **Table 6**).

First, we examined CS programs whose Scholar scores are significantly higher than their USN CS scores. Among the high-ranked CS programs of this type we find Yale University (+0.3), John Hopkins University (+0.5), and UC Santa Cruz (+0.7). Interestingly, these 3 programs are among the 4 smallest CS programs in the top 31 programs ranked by Scholar model. A related subgroup consists of historically solid CS programs such as Colorado State University (+0.6), University of Tennessee (+0.6), College of William and Mary (+0.4), Portland State University (from NR to 2.7), and Lehigh University (+0.6), whose sizes are also small and range from 15 to 24 faculty. We hypothesize that it is more likely that a surveyed peer does not know any faculty in smaller programs and that this might lead to conservative ratings.

A large subgroup of CS programs scoring significantly higher by Scholar model consists of those that have recently experienced significant growth, such as New York University (+0.6; partly due to the recent addition of Tandon School of Engineering), UC Riverside (+0.5), Northeastern University (+0.5), George Mason University (+0.4), UT Dallas (+0.5), UT Arlington (+0.5), Temple University (+0.6), and University of Central Florida (+0.4). This might be explained by the lag between the peer assessments we use in our study (collected in 2009 and 2013) and the citation measures (collected in Fall 2016). In addition, there might be an additional lag between improvements in a program and recognition of the improvements among the peers.

We also examine CS programs that are ranked significantly lower by Scholar model. One possible explanation for some of the lower scores is related to the method we used to identify CS faculty, which includes only faculty listed on people pages of the CS programs. We observe that some universities with highly regarded CS doctoral programs, such as UT Austin (-0.6) and UI Urbana–Champaign (-0.5), have very strong non-CS departments (e.g., Electrical Engineering, Computer Engineering) with a significant number of faculty who publish in the CS venues, but who are not referred at all or are only mentioned as affiliated faculty in people pages of the CS departments. Thus, such faculty are not counted in our study and the resulting cumulative citation scores of those programs are lower than they would be if our inclusion criterion were broader.

## 4 DISCUSSION AND CONCLUSIONS

The main contribution of this work is in showing that there is a high correlation between peer assessments and citation measures of CS doctoral programs. This is a remarkable result considering the subjective nature of peer assessment. In particular, it is likely that each survey responder used different internal metrics when assigning scores and that his or her depth of knowledge about each CS doctoral program varied. Our result demonstrates that committees of imperfect raters could produce good decisions, as has been observed in many other settings [6,19,12].

An open question is whether the correlation between peer assessment and objective measures of quality of CS doctoral programs could be improved further. One way to accomplish this would be to reduce the time gap between peer assessment (in our case, this information comes from 2009 and 2013 surveys) and citation measures (collected during Fall 2016). The correlation could also be improved by addressing some issues with the peer assessments and the objective measures.

On the peer assessment side, one observed issue is that assessment of smaller or less-known programs is noisy and unreliable, as evidenced by the 0.360 correlation measured on a subset of CS programs with the USN CS scores between 2.0 and 2.6. Another issue is a potential influence of the university reputation on peer assessments, with our results indicating that the peer assessment of a CS doctoral program might swing by as much as 0.49 points depending on the reputation of the host university. This demonstrates that peer assessments could be biased by factors indirectly related to the program quality. The root of both issues with the peer assessment is a difficulty in obtaining relevant information about the programs on the side of the assessors. A remedy could be in collecting and publishing unbiased and objective measures of CS doctoral programs to inform the peer assessment.

This study focuses on citation measures of CS doctoral programs. Although we put concerted effort to collect unbiased citation data, there are several issues with the collected data. One issue is related to the definition of a CS doctoral program; does it refer to an administrative unit (e.g., a CS department) with faculty having the primary appointments in that unit, or it refers to all CS-related faculty in the university, regardless of their home department? While our definition is that CS doctoral program refers to the administrative unit, it might be preferable to include all CS-related faculty. In such case, it might be necessary to design a weighting mechanism for faculty that are only partially publishing in the CS venues. Clearly, this approach would be time consuming because it would require searching for CS-related faculty across a whole university.

Our study relies on Google Scholar for the citation data. Despite some limitations of automated web crawling [8,14], the quality of Google Scholar data is comparable to the data coming from the subscription-based services for journal publications such as Web of Science [3]. One advantage of Google Scholar is that it provides coverage of both journal and conference papers, which is very important for a field such as computer science. Google Scholar provides an opportunity for faculty to create and maintain their own publication profile. It is straightforward to extract citation indices of faculty with the profile. Thus, encouraging all CS faculty to have their own profile would make extraction of the citation measures an easier task. While we did not have resources to account for self-citations while calculating the citation indices, this might have been feasible if most CS faculty had Scholar profiles.

In our study, we do not pay attention to the number of authors and position in the author list. Whether and how the order and number of authors in a paper should be factored remains an open question [17]. To calculate the citation measure of a CS doctoral program, we rely on aggregating citation indices of its faculty. Such aggregation does not account for the research field of the faculty. It could be argued that this is unfair to subdisciplines that have smaller communities or require more work to publish a paper [20]. It is also an open question whether the CS doctoral programs that cover a large number of subdisciplines should be ranked higher than the same sized programs that concentrate their faculty in a smaller number of subdisciplines.

This study uses only citation data from Google Scholar to create objective measures of program quality. Numerous other measures have been proposed and used for university and program ranking [1,18]. Related measures that consider publication productivity include counts of papers and counts of papers at top venues. For example, a recently released ranking[10] of CS departments is based on counts of faculty publications in selected CS conferences. Furthermore, this ranking is also making sure the less represented subareas of CS are given a higher weight in ranking and it down-weights papers with many coauthors. Beyond publication data, measures such as faculty recognition, student placement, student selectivity, research funding, resources, and diversity have been proposed. Although our proposed citation measures have high correlation with the peer assessment, it is likely that each additional measure, if collected in an unbiased manner and having sufficient quality, might further improve the explained variance of regression models. Moving forward, it might be helpful to create a central publicly accessible resource where different program measures could be deposited. For such a resource to be truly useful, it would be important to provide raw data, detailed description of data collection process and any potential issues with data quality, such as missing data, bias, and uncertainty, and all the relevant code. In addition to improving regression modeling, this resource would be very useful during peer assessment, by providing peers with objective and unbiased information about the assessed programs.

One caveat with any ranking of universities and programs based on objective measures is that it could be susceptible to gaming [9,11,15]. Thus, to be societally useful, any ranking should be examined for potential negative incentives for a university or a program change or for presence of shortcuts to artificially improve the ranking. Peer assessment informed by the objective measures might play an important role in reducing incentives for gaming of academic rankings. Presence of a publicly accessible resource containing all available objective program measures could further discourage attempts to game the ranking.

[10] http://csrankings.org/

**Table 7.** List of 173 U.S. CS graduate programs: Ranking by our aggregated model (Rank), University name (University), Number of tenured faculty with *t10* score (Size), median *t10* score of all the faculty (M10), geometric mean of t10 score of all faculty (G10), number of highly cited faculty based on *c40* (C40) and *c60* (C60), U.S. News CS score (USN), Scholar score (Scholar)

| Rank | University | Size | M10 | G10 | C40 | C60 | USN | Scholar |
|---|---|---|---|---|---|---|---|---|
| **1** | Carnegie Mellon University | 143 | 218 | 200 | 105 | 74 | 5 | 5 |
| **1** | Massachusetts Institute of Technology | 97 | 306 | 286 | 72 | 66 | 5 | 5 |
| **1** | Stanford University | 55 | 395 | 425 | 46 | 43 | 5 | 5 |
| **1** | University of California - Berkeley | 68 | 375 | 351 | 57 | 54 | 5 | 5 |
| **5** | Cornell University | 75 | 216 | 228 | 50 | 41 | 4.5 | 4.4 |
| **6** | Georgia Institute of Technology | 97 | 167 | 139 | 66 | 48 | 4.3 | 4.3 |
| **6** | University of Washington | 56 | 232 | 239 | 40 | 31 | 4.5 | 4.3 |
| **8** | University of California - Los Angeles | 44 | 206 | 243 | 37 | 28 | 4.1 | 4.2 |
| **8** | University of California - San Diego | 60 | 204 | 192 | 47 | 36 | 4 | 4.2 |
| **10** | Columbia University | 45 | 218 | 206 | 35 | 27 | 4 | 4.1 |
| **10** | Princeton University | 35 | 285 | 232 | 27 | 23 | 4.4 | 4.1 |
| **10** | University of Illinois - Urbana - Champaign | 63 | 169 | 163 | 45 | 34 | 4.6 | 4.1 |
| **10** | University of Michigan - Ann Arbor | 59 | 235 | 175 | 40 | 31 | 4.1 | 4.1 |
| **14** | Johns Hopkins University | 27 | 277 | 296 | 18 | 14 | 3.5 | 4 |
| **14** | New York University | 57 | 204 | 154 | 39 | 31 | 3.4 | 4 |
| **14** | University of Maryland - College Park | 48 | 182 | 174 | 37 | 32 | 4 | 4 |
| **14** | Yale University | 21 | 319 | 261 | 19 | 16 | 3.7 | 4 |
| **18** | University of Southern California | 39 | 234 | 194 | 25 | 20 | 3.7 | 3.9 |
| **18** | University of Wisconsin - Madison | 34 | 215 | 196 | 27 | 22 | 4.2 | 3.9 |
| **20** | California Institute of Technology | 16 | 287 | 238 | 12 | 8 | 4.2 | 3.7 |
| **20** | Harvard University | 30 | 202 | 183 | 22 | 16 | 3.9 | 3.7 |
| **20** | University of Massachusetts - Amherst | 44 | 177 | 170 | 25 | 16 | 3.6 | 3.7 |
| **20** | University of Pennsylvania | 33 | 188 | 165 | 25 | 21 | 3.8 | 3.7 |
| **20** | University of Texas - Austin | 46 | 177 | 142 | 27 | 23 | 4.3 | 3.7 |
| **25** | Duke University | 39 | 163 | 155 | 24 | 17 | 3.6 | 3.6 |
| **25** | University of California - Santa Barbara | 30 | 166 | 158 | 23 | 16 | 3.3 | 3.6 |
| **27** | Brown University | 31 | 167 | 145 | 20 | 17 | 3.7 | 3.5 |
| **27** | University of California - Davis | 33 | 149 | 129 | 23 | 17 | 3.3 | 3.5 |
| **27** | University of California - Santa Cruz | 21 | 192 | 169 | 16 | 13 | 2.8 | 3.5 |
| **27** | University of Chicago | 36 | 144 | 165 | 21 | 15 | 3.3 | 3.5 |
| **27** | University of North Carolina - Chapel Hill | 32 | 161 | 146 | 21 | 16 | 3.6 | 3.5 |
| **32** | Pennsylvania State University | 35 | 136 | 114 | 21 | 15 | 3.4 | 3.4 |
| **32** | University of California - Irvine | 42 | 112 | 116 | 23 | 17 | 3.4 | 3.4 |
| **32** | University of Minnesota - Twin Cities | 44 | 98 | 127 | 29 | 15 | 3.4 | 3.4 |
| **35** | Purdue University | 54 | 112 | 111 | 27 | 14 | 3.7 | 3.3 |
| **35** | Rice University | 19 | 154 | 147 | 15 | 10 | 3.7 | 3.3 |
| **35** | Rutgers University | 41 | 124 | 112 | 20 | 16 | 3.3 | 3.3 |
| **35** | Stony Brook University - SUNY | 42 | 116 | 121 | 22 | 11 | 3.1 | 3.3 |
| **35** | University of California - Riverside | 31 | 134 | 128 | 18 | 12 | 2.8 | 3.3 |
| **40** | Boston University | 24 | 118 | 134 | 16 | 9 | 3 | 3.2 |
| **40** | Northeastern University | 58 | 89 | 99 | 25 | 14 | 2.7 | 3.2 |
| **40** | University of Arizona | 19 | 140 | 130 | 15 | 9 | 3.1 | 3.2 |
| **43** | Northwestern University | 33 | 99 | 104 | 21 | 9 | 3.3 | 3.1 |
| **43** | Ohio State University | 42 | 95 | 90 | 21 | 10 | 3.3 | 3.1 |
| **43** | University of Virginia | 27 | 140 | 81 | 15 | 10 | 3.4 | 3.1 |
| **46** | Colorado State University | 21 | 107 | 110 | 15 | 7 | 2.4 | 3 |
| **46** | Indiana University | 46 | 87 | 80 | 21 | 11 | 2.9 | 3 |
| **46** | Michigan State University | 31 | 96 | 97 | 16 | 7 | 2.8 | 3 |
| **46** | University at Buffalo - SUNY | 37 | 109 | 97 | 13 | 8 | 2.6 | 3 |
| **46** | University of Colorado - Boulder | 34 | 108 | 100 | 15 | 8 | 3.1 | 3 |
| **46** | University of Rochester | 19 | 117 | 124 | 11 | 6 | 2.9 | 3 |
| **46** | University of Tennessee - Knoxville | 24 | 113 | 114 | 12 | 6 | 2.4 | 3 |
| **46** | University of Utah | 43 | 93 | 82 | 20 | 10 | 3.1 | 3 |
| **46** | Virginia Tech | 43 | 79 | 79 | 19 | 10 | 3.1 | 3 |
| **55** | Arizona State University | 59 | 70 | 74 | 19 | 11 | 3 | 2.9 |
| **55** | George Mason University | 41 | 80 | 69 | 20 | 7 | 2.5 | 2.9 |
| **55** | North Carolina State University | 49 | 62 | 63 | 18 | 11 | 3 | 2.9 |
| **55** | Texas A&M University | 42 | 76 | 74 | 21 | 8 | 3.1 | 2.9 |
| **55** | University of Texas - Dallas | 50 | 72 | 54 | 23 | 10 | 2.4 | 2.9 |
| **55** | Vanderbilt University | 21 | 113 | 110 | 12 | 5 | 2.8 | 2.9 |
| **55** | Washington University | 25 | 96 | 112 | 13 | 6 | 3.1 | 2.9 |
| **62** | College of William and Mary | 15 | 118 | 94 | 8 | 4 | 2.4 | 2.8 |
| **62** | Rensselaer Polytechnic Institute | 24 | 80 | 91 | 11 | 6 | 2.9 | 2.8 |
| **62** | University of Pittsburgh | 18 | 109 | 102 | 10 | 4 | 2.9 | 2.8 |
| **65** | Dartmouth College | 21 | 87 | 96 | 10 | 3 | 3.1 | 2.7 |
| **65** | Lehigh University | 16 | 73 | 94 | 9 | 4 | 2.1 | 2.7 |
| **65** | Portland State University | 22 | 90 | 56 | 10 | 7 | 0 | 2.7 |
| **65** | University of Florida | 41 | 51 | 61 | 12 | 8 | 3 | 2.7 |
| **65** | University of Illinois - Chicago | 31 | 66 | 74 | 10 | 6 | 2.7 | 2.7 |
| **65** | University of Maryland - Baltimore County | 25 | 71 | 88 | 10 | 5 | 2.4 | 2.7 |
| **65** | University of Notre Dame | 19 | 95 | 86 | 9 | 4 | 2.7 | 2.7 |
| **65** | University of Texas - Arlington | 26 | 74 | 63 | 12 | 5 | 2.2 | 2.7 |
| **73** | CUNY Grad School & University Center | 78 | 29 | 30 | 22 | 10 | 2.3 | 2.6 |
| **73** | Temple University | 22 | 72 | 60 | 10 | 5 | 2 | 2.6 |
| **73** | University of Central Florida | 36 | 53 | 58 | 11 | 8 | 2.2 | 2.6 |
| **73** | University of Nebraska - Lincoln | 28 | 71 | 65 | 10 | 5 | 2.4 | 2.6 |
| **77** | Illinois Institute of Technology | 18 | 65 | 57 | 8 | 3 | 2.1 | 2.5 |
| **77** | University of Delaware | 24 | 67 | 63 | 8 | 3 | 2.4 | 2.5 |
| **77** | University of South Florida | 21 | 63 | 59 | 8 | 5 | 2.1 | 2.5 |
| **80** | Case Western Reserve University | 18 | 71 | 62 | 6 | 2 | 2.4 | 2.4 |
| **80** | Drexel University | 19 | 54 | 65 | 6 | 2 | 2.2 | 2.4 |
| **80** | New Jersey Institute of Technology | 29 | 49 | 41 | 9 | 4 | 2.2 | 2.4 |
| **80** | Tufts University | 16 | 62 | 64 | 5 | 2 | 2.4 | 2.4 |
| **80** | University of Houston | 24 | 60 | 55 | 11 | 1 | 2.1 | 2.4 |
| **80** | University of Memphis | 11 | 72 | 64 | 4 | 2 | 0 | 2.4 |
| **80** | University of Missouri | 15 | 66 | 56 | 6 | 3 | 2.1 | 2.4 |
| **80** | University of New Mexico | 17 | 55 | 65 | 6 | 2 | 2.3 | 2.4 |
| **80** | Wayne State University | 20 | 65 | 69 | 7 | 1 | 2 | 2.4 |
| **80** | Worcester Polytechnic Institute | 27 | 57 | 56 | 7 | 3 | 2.2 | 2.4 |
| **90** | Brandeis University | 14 | 59 | 34 | 6 | 4 | 2.3 | 2.3 |
| **90** | Brigham Young University | 31 | 50 | 39 | 7 | 3 | 2.2 | 2.3 |
| **90** | Georgia State University | 19 | 45 | 50 | 6 | 2 | 2 | 2.3 |
| **90** | Oregon State University | 36 | 57 | 47 | 8 | 1 | 2.5 | 2.3 |
| **90** | University of Connecticut | 24 | 60 | 50 | 5 | 1 | 2.3 | 2.3 |
| **90** | University of Iowa | 20 | 56 | 59 | 5 | 2 | 2.6 | 2.3 |
| **90** | University of Kansas | 23 | 46 | 47 | 5 | 3 | 2.3 | 2.3 |
| **90** | University of Texas - San Antonio | 24 | 58 | 43 | 6 | 1 | 0 | 2.3 |
| **90** | West Virginia University | 21 | 61 | 39 | 8 | 2 | 2 | 2.3 |
| **99** | Binghamton University - SUNY | 27 | 51 | 44 | 4 | 1 | 2 | 2.2 |
| **99** | Clemson University | 33 | 54 | 42 | 5 | 1 | 2.3 | 2.2 |
| **99** | Florida International University | 29 | 51 | 27 | 6 | 2 | 0 | 2.2 |
| **99** | Iowa State University | 27 | 42 | 45 | 5 | 2 | 2.6 | 2.2 |
| **99** | Louisiana State University - Baton Rouge | 19 | 47 | 38 | 3 | 2 | 2.1 | 2.2 |
| **99** | Missouri University of Science & Tech | 16 | 52 | 43 | 4 | 1 | 2 | 2.2 |
| **99** | Syracuse University | 26 | 38 | 34 | 6 | 3 | 2.5 | 2.2 |
| **99** | University at Albany - SUNY | 18 | 58 | 49 | 4 | 1 | 2.1 | 2.2 |
| **99** | University of Georgia | 21 | 51 | 48 | 4 | 2 | 2.2 | 2.2 |
| **99** | University of Hawaii - Manoa | 24 | 36 | 27 | 6 | 3 | 0 | 2.2 |
| **99** | University of Kentucky | 23 | 45 | 32 | 4 | 3 | 2.2 | 2.2 |
| **99** | University of Oregon | 15 | 43 | 58 | 3 | 1 | 2.6 | 2.2 |
| **99** | University of South Carolina | 25 | 33 | 42 | 7 | 1 | 2.1 | 2.2 |
| **112** | Florida Atlantic University | 31 | 26 | 26 | 6 | 3 | 0 | 2.1 |
| **112** | Florida State University | 22 | 44 | 37 | 5 | 1 | 2.3 | 2.1 |
| **112** | George Washington University | 15 | 32 | 33 | 3 | 2 | 2.3 | 2.1 |
| **112** | Stevens Institute of Technology | 14 | 45 | 44 | 3 | 1 | 2.1 | 2.1 |
| **112** | University of Massachusetts - Lowell | 15 | 42 | 23 | 4 | 2 | 0 | 2.1 |
| **117** | Colorado School of Mines | 9 | 58 | 48 | 2 | 0 | 2.1 | 2 |
| **117** | DePaul University | 50 | 19 | 22 | 6 | 3 | 0 | 2 |
| **117** | Naval Postgraduate School | 18 | 33 | 32 | 2 | 1 | 2.4 | 2 |
| **117** | Old Dominion University | 15 | 39 | 28 | 2 | 1 | 0 | 2 |
| **117** | University of Alabama | 16 | 25 | 25 | 4 | 2 | 0 | 2 |
| **117** | University of Alabama - Birmingham | 9 | 37 | 40 | 1 | 1 | 0 | 2 |
| **117** | University of Arkansas - Fayetteville | 16 | 32 | 22 | 2 | 2 | 0 | 2 |
| **117** | University of Maine | 6 | 50 | 34 | 3 | 0 | 0 | 2 |
| **117** | University of Massachusetts - Boston | 15 | 41 | 43 | 5 | 0 | 0 | 2 |
| **117** | University of Tulsa | 13 | 32 | 35 | 3 | 1 | 0 | 2 |
| **117** | Virginia Commonwealth University | 17 | 36 | 37 | 2 | 1 | 0 | 2 |
| **117** | Washington State University | 17 | 34 | 20 | 4 | 1 | 2.4 | 2 |
| **129** | Claremont Graduate University | 6 | 38 | 45 | 1 | 0 | 0 | 1.9 |
| **129** | Kansas State University | 15 | 43 | 32 | 3 | 0 | 2.2 | 1.9 |
| **129** | New Mexico State University | 13 | 56 | 31 | 1 | 0 | 0 | 1.9 |
| **129** | Oakland University | 14 | 30 | 28 | 1 | 1 | 0 | 1.9 |
| **129** | University of Colorado - Colorado Springs | 12 | 21 | 27 | 2 | 1 | 0 | 1.9 |
| **129** | University of Nevada - Reno | 18 | 21 | 29 | 2 | 1 | 0 | 1.9 |
| **129** | University of North Carolina - Charlotte | 20 | 39 | 29 | 3 | 0 | 2.1 | 1.9 |
| **129** | University of Oklahoma | 15 | 38 | 39 | 2 | 0 | 2 | 1.9 |
| **129** | University of Texas - El Paso | 15 | 25 | 25 | 2 | 1 | 0 | 1.9 |
| **138** | Oregon Health and Science University | 8 | 35 | 37 | 1 | 0 | 2.2 | 1.8 |
| **138** | University of Cincinnati | 20 | 28 | 24 | 1 | 1 | 2 | 1.8 |
| **138** | University of Denver | 8 | 47 | 29 | 1 | 0 | 0 | 1.8 |
| **138** | University of Louisiana - Lafayette | 22 | 15 | 23 | 3 | 1 | 0 | 1.8 |
| **138** | University of Louisville | 15 | 21 | 14 | 3 | 1 | 0 | 1.8 |
| **138** | University of North Texas | 27 | 37 | 26 | 2 | 0 | 0 | 1.8 |
| **138** | University of Wisconsin - Milwaukee | 12 | 36 | 32 | 1 | 0 | 0 | 1.8 |
| **138** | Utah State University | 14 | 30 | 24 | 3 | 0 | 0 | 1.8 |
| **146** | Auburn University | 17 | 22 | 21 | 1 | 0 | 2.2 | 1.7 |
| **146** | Florida Institute of Technology | 20 | 17 | 13 | 1 | 1 | 0 | 1.7 |
| **146** | Kent State University | 19 | 11 | 11 | 2 | 1 | 0 | 1.7 |
| **146** | Texas Tech University | 17 | 14 | 20 | 1 | 1 | 0 | 1.7 |
| **146** | University of Alabama - Huntsville | 16 | 24 | 24 | 1 | 0 | 0 | 1.7 |
| **146** | University of Arkansas - Little Rock | 8 | 14 | 16 | 1 | 1 | 0 | 1.7 |
| **146** | University of New Orleans | 11 | 26 | 20 | 1 | 0 | 0 | 1.7 |
| **153** | Mississippi State University | 17 | 18 | 19 | 1 | 0 | 0 | 1.6 |
| **153** | Montana State University | 9 | 30 | 28 | 0 | 0 | 0 | 1.6 |
| **153** | Southern Methodist University | 12 | 16 | 12 | 2 | 0 | 2 | 1.6 |
| **153** | University of Mississippi | 8 | 18 | 11 | 1 | 0 | 0 | 1.6 |
| **153** | University of Missouri - Kansas City | 19 | 22 | 17 | 1 | 0 | 0 | 1.6 |
| **153** | University of Wyoming | 6 | 20 | 19 | 1 | 0 | 0 | 1.6 |
| **153** | Western Michigan University | 13 | 19 | 11 | 1 | 0 | 0 | 1.6 |
| **160** | Air Force Institute of Technology | 12 | 19 | 19 | 0 | 0 | 0 | 1.5 |
| **160** | Michigan Technological University | 16 | 18 | 18 | 0 | 0 | 0 | 1.5 |
| **160** | North Dakota State University | 7 | 18 | 15 | 0 | 0 | 0 | 1.5 |
| **160** | Towson University | 28 | 10 | 12 | 1 | 0 | 0 | 1.5 |
| **160** | University of Idaho | 13 | 19 | 11 | 0 | 0 | 0 | 1.5 |
| **160** | University of Southern Mississippi | 13 | 7 | 10 | 2 | 0 | 0 | 1.5 |
| **166** | New Mexico Institute of Mining & Tech | 7 | 11 | 11 | 0 | 0 | 0 | 1.4 |
| **166** | Nova Southeastern University | 26 | 1 | 5 | 1 | 1 | 0 | 1.4 |
| **166** | Oklahoma State University | 12 | 17 | 12 | 0 | 0 | 0 | 1.4 |
| **166** | University of Colorado - Denver | 10 | 8 | 12 | 0 | 0 | 0 | 1.4 |
| **166** | University of Nebraska - Omaha | 17 | 13 | 9 | 0 | 0 | 0 | 1.4 |
| **171** | Louisiana Tech University | 6 | 7 | 5 | 0 | 0 | 0 | 1.3 |
| **172** | Indiana State University | 7 | 0 | 2 | 0 | 0 | 0 | 1.1 |
| **172** | LIU Post | 4 | 0 | 1 | 0 | 0 | 0 | 1.1 |